\documentstyle[aps,preprint,epsfig]{revtex}
\begin{document}

\title{Hadronic interactions models beyond collider energies}
\author{L. A. Anchordoqui, M. T. Dova, L. N. Epele and S. J. Sciutto}
\address{Departamento de F\'{\i}sica, Universidad Nacional de La
Plata \\
C.C. 67, (1900) La Plata \\ Argentina}
\maketitle

\begin{abstract}

Studies of the influence of different hadronic models on extensive air
showers at ultra-high energies are presented. The hadronic models
considered are those implemented in the well-known QGSJET and SIBYLL
event generators. The different approaches used in both codes to model
the underlying physics is analyzed using computer simulations
performed with the program AIRES. The most relevant observables for
both single collisions and air showers are studied for primary
energies ranging from $10^{14}$ eV up to $10^{20.5}$ eV. In addition,
the evolution of lateral and energy distributions during the shower
development is presented.
Our analysis seems to indicate that the behaviour of shower
observables does not largely reflect the strong differences observed in
single collisions.
\hfill

\noindent {\it PACS number(s):} 96.40.Tv, 13.85.Tp, 96.40Pq.
\end{abstract}

\newpage
\section{Introduction}

Extremely high energy cosmic rays (CR) are an extravagant
phenomenon of nature that has baffled astrophysics for more than three
decades \cite{Linsley}.
Ingenious installations with large effective area and long exposure
times to overcome the rapidly decreasing flux, $\sim$ 1 event per km$^2$
per year (century) at $10^{19}$ ($10^{20}$) eV, are required to study
them. Their
energy spectrum beyond
1 PeV
needs to be studied indirectly through the extensive air
showers (EAS) they produce deep in the atmosphere. Thus, the
interpretation of the observed cascades generally depends on  Monte Carlo
simulations which extrapolate hadronic interaction models to energies
well beyond those explored at acellerators.

There is a couple of quite elaborate models (the dual parton model
(DPM) \cite{dpm}
and the quark gluon string (QGS) model of the supercritical
Pomeron \cite{qgs})
that provide a complete
phenomenological
description of all facets of soft hadronic processes. These models, inspired
on $1/N$ expansion of QCD are also supplemented with
generally
accepted theoretical principles like duality, unitarity, Regge behavior and
parton structure. At higher energies, however, there is evidence of minijet
production \cite{UA1} and correlation between
multiplicity per event and transverse momentum per particle \cite{ua1},
suggesting that
semihard QCD processes become important in high energy hadronic
interactions. It is precisely the problem of a proper accounting for semihard
processes the major source of uncertainty of extensive air showers event
generators.

Two codes of hadronic interactions with similar
underlying physical assumptions and algorithms tailored for
efficient operation to the highest cosmic ray energies are
SIBYLL \cite{sybill} and QGSJET \cite{qgsjet}.
In these codes, the low $p_{\rm T}$ interactions are modeling by the
exchange of
Pomerons. Regge singularities are used to determine the momentum
distribution functions of the various sets of constituents, valence
and sea quarks. In the interaction the hadrons exchange very soft
gluons simulated by the production of a single pair QCD strings and
the subsequent fragmentation into colour neutral hadrons. In QGSJET 
these events
also involve exchange of multiple pairs of soft strings.

As above mentioned, the production of small jets is expected to
dominate interactions in the c.m. energy above $\sqrt{s}\approx$ 40 TeV.
The underlying idea behind SIBYLL is that the increase in
the cross section is driven by the production of
minijets \cite{minijet}. The probability distribution for obtaining $N$
jet pairs (with $p_{\rm T}^{\rm jet}\,>\,p_{\rm T}^{\rm min}$, being
$p_{\rm T}^{\rm min}$ a sharp threshold on the transverse momentum
below which hard interactions are neglected)
in a collision  at energy $\sqrt{s}$ is computed regarding elastic $pp$ or
$p\bar{p}$ scattering as a difractive shadow scattering associated
with inelastic processes \cite{durandpi}. The algorithms are
tuned to reproduce the central and fragmentation regions data
up to $p\bar{p}$ collider energies,
and
with no further adjustments they are extrapolated several orders of magnitude.

In QGSJET the theory is formulated entirely in terms of Pomeron
exchanges. The basic idea is to replace the soft Pomeron by a
so-called
``semihard Pomeron'', which is defined to be an ordinary soft Pomeron
with the middle piece replaced by a QCD parton ladder.
Thus, minijets will emerge as a part of the ``semihard Pomeron'',
which is itself the controlling mechanism for the whole interaction.
After
performing the energy sharing among the soft and semihard Pomerons, and
also the sharing
among the soft and hard pieces of the last one; the number of charged
particles
in the partonic cascade is easily obtained generalizing the method of
multiple production of hadrons
as discussed in the QGS model (soft Pomeron showers) \cite{qgs}.

The most outstanding point
in connection with the shower development is certainly the incorporation
of nuclear effects.
Both, SIBYLL and QGSJET \cite{kaida} describe particle production in
hadron-nucleus
collisions in a quite similar fashion. The high energy projectile undergoes
a multiple scattering as formulated in Glauber's approach \cite{glauber},
particle
production comes again after the fragmentation of colorless parton-parton
chains
constructed from the quark content of the interacting hadrons.
In cases with more than one wounded nucleon in the target the extra strings
are connected with sea-quarks in the projectile.
This
ensures that the inelasticity in hadron-nucleus collisions is not much
larger than that corresponding to hadron-hadron collisions.
A higher inelastic nuclear stopping power
yields relatively rapid shower developments which are
ruled out by $p$-nucleus data \cite{frichter}.

Except by the depth of maximum of the shower -- a quantity
which is well known to depend on the location and features of the
first interaction -- one can expect that global observables should not
be affected by the above reasonable alternative physics assumptions.
However, the
question on the sensitivity of the free parameters of these models
(which have been derived from available accelerator data) when they
are extrapolated to energies essentially
greater than attained with colliders, is
surely an interesting one. To answer this question is the main goal of
the present article.

In this work we shall present several comparative studies between
SIBYLL and QGSJET \cite{Knapp}. The outline of the paper is as follows.
In Sec. II we
proceed by first analyzing the different predictions on  $p$-air
and $\pi$-air cross sections, and then we compare single $p$-$\bar{p}$ and
$\bar{p}$-nuclei hadronic interactions.  In Sec. III we present results
of several numerical analyses. Around 5000
air showers induced by protons with energies ranging from $10^{14}$ up
to $10^{20.5}$ eV are generated
with the code AIRES \cite{sergio}, a realistic air shower simulation 
system which includes
electromagnetic interactions algorithms \cite{hillas} and links to the
mentioned SIBYLL and QGSJET models. We conclude in Sec. IV with the final
remarks.

\newpage

\section{Hadronic collisions}

Let us start our comparative analysis of SIBYLL and QGSJET by discussing
briefly the $p$-air and $\pi$-air cross sections as calculated by these
models and also by the simulation program AIRES.

In Fig. 1 (2) the $p$-air ($\pi$-air) cross section is plotted versus the
projectile laboratory energy. In the cases of AIRES cross sections (solid
lines) --which are equivalent to the so-called ``Bartol cross sections''--
and
QGSJET (dotted lines), the mentioned laboratory energy
is the input energy for
the corresponding algorithms. On the other hand, the input parameter for
SIBYLL  is the c.m. energy in the hadron-{\em nucleon\/}
system \cite{Tom}. Notice that similar plots can be found in Ref.
\cite{qgsjet} with a different behaviour of the SIBYLL cross sections. We
attribute these differences to the c.m. conversion procedure. In
fact, if the laboratory energy is (mistakenly) converted to the
hadron-{\em nucleus\/} system, one may reproduce the data of Ref.
\cite{qgsjet}.

Note that in QGSJET the growth in the cross section is fitted using a
mixture of soft and hard interactions,
contrariwise, SIBYLL does so just with the hard processes (minijets).
Thus, it is feasible
to expect
a deviation in the predictions when the algorithms are extrapolated several
orders of magnitude. Furthermore, SIBYLL predictions for the $p$-$\bar{p}$
cross section ought to be higher than the ones of QGSJET since
hard processes overrule the soft ones with the rise of energy.
As aforementioned, the extension to hadron nuclei interactions is computed
in both codes in the framework of Glauber theory with minute differences,
yielding no significant additional divergences. As a consequence, we
attribute the different behaviours for the cross sections shown in Figs.
1 and 2, to the way in which the free parameters of both codes are fitted to
reproduce $p$-$\bar{p}$ collider data. Namely, $p^2_{\rm min} = 5$ GeV$^2$
and the multiplicative {\em ad hoc} factor $k=1.7$ in SIBYLL
code.\footnote{Actually
one still has another parameter but hardly would have any influence at
high energies. Recall that in SIBYLL
the soft part of
the eikonal function is taken as a constant fitted to low energy data
\cite{sybill}.} On the other hand, in QGSJET we have: i)
parameters of the Pomeron trajectory:
$\Delta = 0.07$, $\alpha'_{\rm P}(0) = 0.21$ GeV$^{-2}$, ii) the ones from
the Pomeron vertices: $R^2_{pp} = 3.56$ GeV$^{-2}$,
$\gamma_p^2$ = 3.64 GeV$^{-2}$, iii) the so-called ``shower
enhancements coefficient $C_{pp} = 1.5$, and iv) the parameters of
semihard
processes: $p_{\rm min}^2$ = 4 GeV$^2$ (this parameter as in the SIBYLL
code
represent the threshold of hard interactions), the parameter associated
with the parton density $r^2=0.6$ GeV$^{-2}$; for a survey the reader is
referred again to \cite{qgsjet} and references therein. It is interesting 
to remark
that with an alternative value in the Pomeron trajectory parameter one can
reproduce the cross section without hard processes, {\em viz}. with QGS
model (see Table I of \cite{qgsjet}).

The most direct way to analyze the differences between the models is to
study the characteristic of the secondaries
generated under similar conditions. For each
hadronic code we generate sets of $10^5$ collisions in order
to analyze the secondaries produced by SIBYLL and QGSJET in $\bar{p}p$ and
$\bar{p}A$ ($A$ represents a nucleus target of mass number $A=10$)
at different projectile energies. 

In all the considered cases we found that the number of secondaries
coming from QGSJET collisions is larger than the ones corresponding
to the SIBYLL case. This shows up clearly in the energy versus
total number of secondaries two-dimension distributions.
Although differences are not quite obvious at
100 TeV (Figs. 3 and 4) they grow up
dramatically when the
projectile energy reaches $10^{20}$ eV (Figs. 5
and 6). With the exception
of protons and neutrons, the normalized energy
distributions possess similar shapes for both models.

Again we remark that the case of $p$-nuclei collisions do not show major
differences with respect to $p$-$\bar{p}$ case.

\newpage

\section{Air shower simulations}

Proton induced air showers are generated using AIRES+SIBYLL and
AIRES+QGSJET. Primary energies range from $10^{14}$ eV up to $10^{20.5}$
eV. To put into evidence as much as possible the differences between
the intrinsic mechanism of SIBYLL and QGSJET we have always used the
same cross sections for hadronic collisions, namely, the AIRES cross
section
plotted in Figs. 1 and 2.

All hadronic collisions with projectile energies below
200 GeV are
processed with the Hillas Splitting algorithm \cite{hillas}, and the
external
collision package is invoked for all those collisions with energies
above the mentioned threshold. It is worthwhile mentioning that for
ultra-high energy primaries, the
low energy collisions represent a little fraction (no more than 10\% at
 $10^{20.5}$ eV)
of the total number of inelastic hadronic processes that take place
during the shower development.
It is also important to stress that the dependence of the shower
observables on the hadronic model is primarily related to the first
interactions which in all the cases are ultra high energy processes
involving only the external hadronic models.
All shower particles with energies
above the following thresholds were
tracked: 500 keV for gammas, 700 keV for electrons and positrons, 1
MeV for muons, 1.5 MeV for mesons and 80 MeV for nucleons and nuclei.
The particles were injected at the top of the atmosphere (100
km.a.s.l) and the ground level was located at sea level.

We have analyzed in detail the longitudinal development of the showers.
The number and energy of different kind of particles have been recorded
as a function of the vertical depth for a number of different observing
levels (more than 100).

The charged multiplicity, essentially electrons and positrons,
is used to determine the number of particles and the location
of the shower maximum by means of four-parameter fits to the
Gaisser-Hillas function \cite{sergio}.

In Fig. 7 $\langle X_{\rm max}\rangle$ is plotted versus
the logarithm of the primary energy for both the SIBYLL and QGSJET
cases. It shows up clearly that SIBYLL showers present higher values
for the depth of the maximum, and that the differences between the
SIBYLL and QGSJET cases increase with the primary energy.
This is consistent with the fact that SIBYLL produces less secondaries
than QGSJET --as discussed in the previous section-- and as a result,
there is a delay in the electromagnetic shower development which is
strongly correlated with $\pi^0$ decays.
The fluctuations, represented by the error bars, decrease
monotonously as long as the energy increases, passing roughly from 95
g/cm$^2$ at $E=10^{14}$ eV to 70 g/cm$^2$ at $E=10^{20.5}$ eV. \footnote{
At this stage it must be stressed that the mean values
we have obtained for depths of shower maximum are slightly different
to those recently presented
by Pryke and Voyvodick \cite{pryke}. The main difference
arise from the fact that
in our treatment the mean free paths for hadron-nucleus collisions
are the same in both models.
Besides, one cannot say that interactions lengths are quite similar
in these two models at $10^{19}$ eV (see Figs. 1 and 2),
unless both are seen in the framework of QGSJET (see discussion in
Sec. II).}

As a representative case we are going to consider in more detail
the behaviour of $10^{20.5}$ eV proton showers. In Fig. 8 the total number
of pions, muons, gammas, and
charged particles are plotted versus the
vertical atmospheric depth. The observed behaviour of the
electromagnetic shower is consistent with the discussion of the previous
paragraphs: A shift in the maximum of the shower. It is worth to mention
that even if the longitudinal development shows important differences at
first stages, they decrease monotonously as far as the shower evolves.
The smallest difference between models corresponds to the case of pions
whose number at the ground level is very similar in both SIBYLL and QGSJET
cases. On the other hand, the number of ground muons does present significant 
(even if not critical) differences.

With the particle data recorded we have evaluated lateral and energy
distributions not only at ground altitude but also {\em at
predetermined observing levels}. To the best of our knowledge this is
the first time that the evolution of lateral and energy distributions
along the longitudinal shower path is studied in such detail. In this
paper we present the distributions corresponding to a subset of all
the levels considered, taking into account particles whose distances
to the shower axis are larger than 50 m.

The high-altitude lateral distributions (Figs. 9, 10 and 11) show
important differences between SIBYLL and QGSJET; such differences
diminish as long as the shower front gets closer to the ground
level. The behaviour can be explained taking into account the
differences between the number of SIBYLL and QGSJET secondaries
reported in the previous section. Due to the fact that SIBYLL produces
less number of secondaries, they have --in average-- more energy and
therefore the number of generations of particles undergoing hadronic
collisions is increased with respect to the QGSJET case. As a result,
during the shower development SIBYLL is called more times than QGSJET,
and this generates a compensation that tends to reduce the difference
in the {\em final\/} number of hadronic secondaries produced during
the entire shower, and consequently in the final decay products, that
is, electrons, gammas and muons.

The lateral distributions of electromagnetic particles are remarkably
similar at both $\langle X_{\rm max}\rangle$
and ground level.\footnote{We want to
stress that $\langle X_{\rm max}\rangle$ is different in each model.} 
However, it comes out from a more detailed analysis of the ground
distributions that they are not strictly coincident and that
the ratio between SIBYLL and QGSJET predictions does depend on $r$,
the distance from the core. In fact, for electrons, this ratio runs
from 1.25 for small $r$ to 0.73 for $r\sim 1000$ m, being equal to 1
at $r \sim 350$ m. A similar behaviour is observed for gammas where
the lateral distributions intersect at $r\sim 1000$ m.

In the case of lateral muon distributions, QGSJET predicts a higher
density for all distances, but the SIBYLL/QGSJET ratio is not
constant, ranging from 0.74 near the core to 0.56 at 1000 m.

The energy distributions at varying altitudes (Figs. 12, 13 and 14)
permit following the dynamics of the shower in great detail. The plots
corresponding to 200 g/cm$^2$ clearly indicate that at such altitude
there is an important fraction of high energy particles (more than 1
TeV). In particular, this is more evident for electrons and gammas
(Figs. 12 and 14, compare the 200 g/cm$^2$ with the respective sea
level distributions).

Finally, analyzing in detail the energy distributions of muons at
ground level, we observe that the ratio of $dN/d\log E$ between SIBYLL
and QGSJET is not constant: At the low (high) end of the spectrum it
takes the value 1.0 (0.8) reaching a minimum of 0.54 around 250 GeV.

\newpage

\section{Conclusions}

Addressing the theoretical issues surrounding high energy hadronic collisions
is intrinsically complicated since many variables are involved. However,
this is crucial in understanding the data  being recorded
by present extremely high energy CR experiments (like the Akeno Giant Air
Shower Array \cite{agasa}) as well as CR next generation experiments
(the future Pierre Auger
Observatory \cite{auger} --fluorescence detector plus ground array--
and the ``eyes'' of
the OWL \cite{owl} that will deeply watch into the CR-sky). 

In this work we have studied the sensitivity of parameters of
 hadronic interactions
models (fitted to low energy data) when the algorithms are
extrapolated several order of magnitudes.  Perhaps the most
oustanding difference between SIBYLL and QGSJET, as we had expected
from our theoretical analysis, is reflected in the predicted number of
secondaries after single $p$-$\bar{p}$ and $\bar{p}$-nuclei
collisions. Such a difference increase steeply with rising
energy. Our investigation on air showers throws up various other
points of interest. In particular, we have reported that the different
number of secondaries predicted remains noticeable during the first
stages of the shower development. It is, of course, immediately
evident that this follows again as a direct consequence of the lower
inelasticity implemented in the SIBYLL generator when compared with
the one in QGSJET.  The study of the evolution of lateral and energy
distributions along the longitudinal shower path allows us to clearly
observe how the differences in the distributions become monotonously
damped, yielding rather similar shapes when reaching the ground.
Further, we have shown that the differences observed at ground level
do depend on the distance to the shower core. Consequently, we are
convinced that it will be possible to obtain relevant information
about the hadronic interactions in air showers from the measurement of
particle densities at distances far from as well as close to the
shower core. This can be achieved if CR experiments are designed
with appropiate dynamic ranges.

On the other hand, in our opinion most of the model discrepancies
discussed in Sec. II will be naturally reduced with the help of data
obtained from future accelerator experiments like the well known Large
Hadron Collider (LHC).

\acknowledgments
It is a pleasure to thank Tom Gaisser for valuable comments on technical
aspects of SIBYLL.
We are also indebted to Alberto Etchegoyen for granting us
computing facilities. This work has been partially supported by
CONICET and the FOMEC program.

\newpage

\section{Legends}

\noindent{\bf Fig. 1:} In the figure we have plotted the corresponding $p$-air
cross sections of SIBYLL (dashed line), QGSJET (dots), and  AIRES
(solid line). We have also superimposed experimental data obtained from
collider experiments $\ast$\cite{carroll}, $\diamondsuit$\cite{roberts},
together with the ones obtained in cosmic ray experiments $\circ$\cite{hara},
$\bullet$\cite{honda}, $\Box$\cite{baltrusaitis}.

\noindent{\bf Fig. 2:} The corresponding $\pi$-air cross sections of SIBYLL, QGSJET
and AIRES. The conventions adopted are the ones of Fig. 1.

\noindent{\bf Fig. 3:} The figure displays the two--dimension distributions (energy vs
number of secondaries) obtained
from $p$-$\bar{p}$ scatterings (incident energy 100 TeV).
In the left hand side we present
the results of QGSJET while the right hand side corresponds to
the ones of SIBYLL. 

\noindent{\bf Fig. 4:} $p$-nuclei ($A$=10) scatterings with incident energies of 100 TeV.
We have used the conventions of Fig. 3.

\noindent{\bf Fig. 5:} $p$-$\bar{p}$ scatterings with incident energies of 100 EeV.
The conventions are the same as Fig. 3.

\noindent{\bf Fig. 6:} $p$-nuclei ($A$=10) scatterings with incident energies of 100 EeV.
We have used the conventions of Fig. 3.

\noindent{\bf Fig. 7:} Simulation results for the average slant depth of maximum, $\langle
X_{\rm max}\rangle$, for proton induced showers, plotted versus the
logarithm of the primary energy. The error bars indicate the standard
fluctuations (the RMS fluctuations of the means are always smaller
than the symbols). The squares (circles) correspond to SIBYLL (QGSJET).

\noindent{\bf Fig. 8:} Longitudinal development of $10^{20.5}$ vertical
proton showers. The average numbers of
particles are plotted versus the atmospheric depth. The solid (dashed)
line stands for the QGSJET (SIBYLL) case.

\noindent{\bf Fig. 9:} Comparison between the recorded electron lateral distributions
displayed by
SIBYLL (grey) and QGSJET (black) at different atmospheric altitudes
including the depth where the shower developes its maximum and the 
predictions at the ground level.

\noindent{\bf Fig. 10:} Same as Fig. 9 for the case of muons.

\noindent{\bf Fig. 11:} Same as Fig. 9 for the case of gammas.

\noindent{\bf Fig. 12:} Electron energy distributions obtained with 
SIBYLL (grey) and QGSJET (black) at different atmospheric altitudes, see level
and depth of shower maximum.

\noindent{\bf Fig. 13:} Same as Fig. 12 for the case of muons.

\noindent{\bf Fig. 14:} Same as Fig. 12 for the case of gammas.

\newpage

\begin{figure}[htbp]
\label{pair}
\begin{center}
\epsfig{file=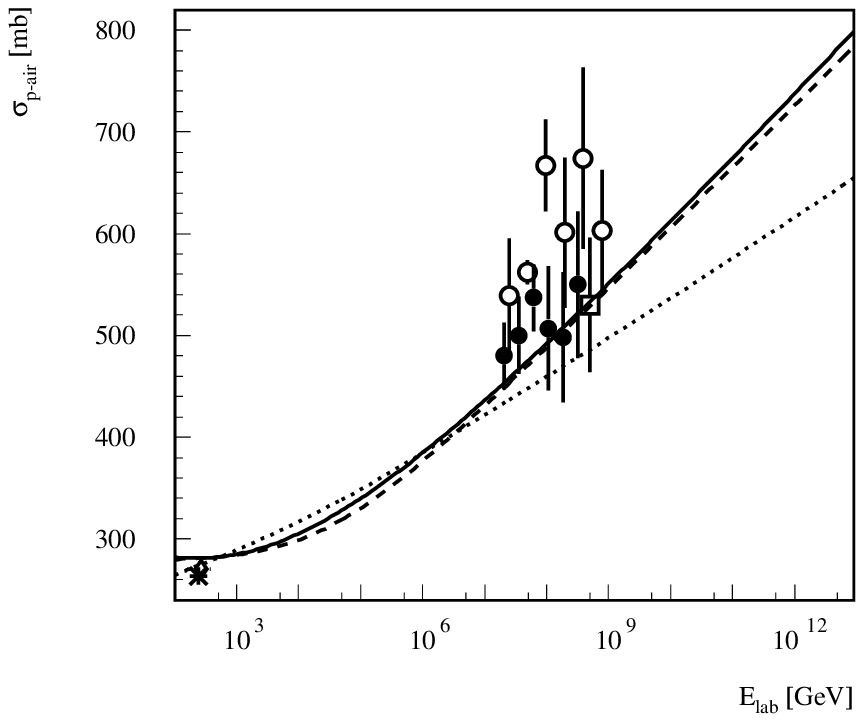,width=15cm,clip=}
\caption{}
\end{center}
\end{figure}

\newpage

\begin{figure}[htbp]
\label{piair}
\begin{center}
\epsfig{file=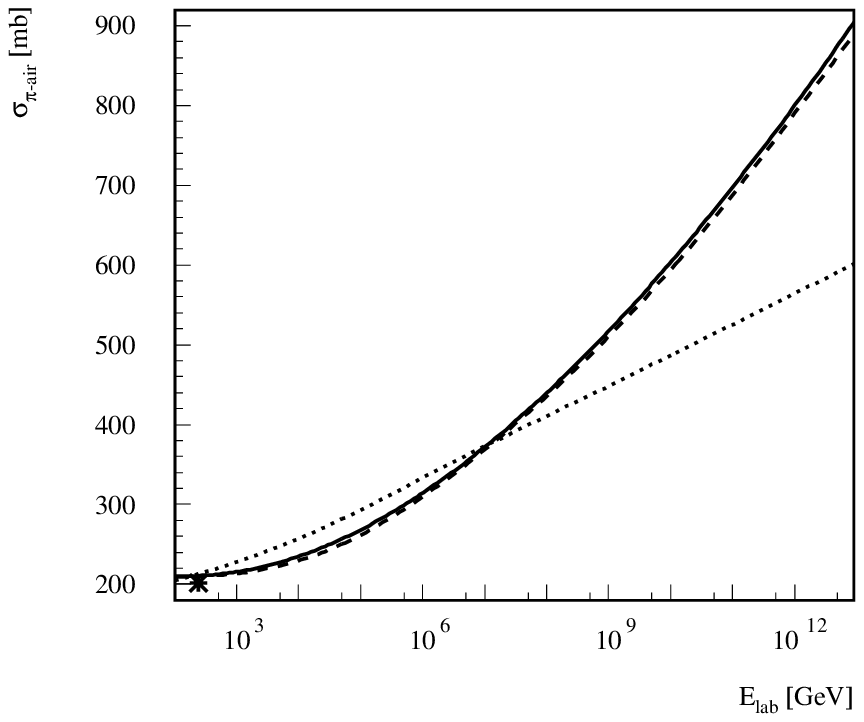,width=15cm,clip=}
\caption{}
\end{center}
\end{figure}

\newpage

\begin{figure}[htpb]
\label{pacota1}
\begin{center}
\epsfig{file=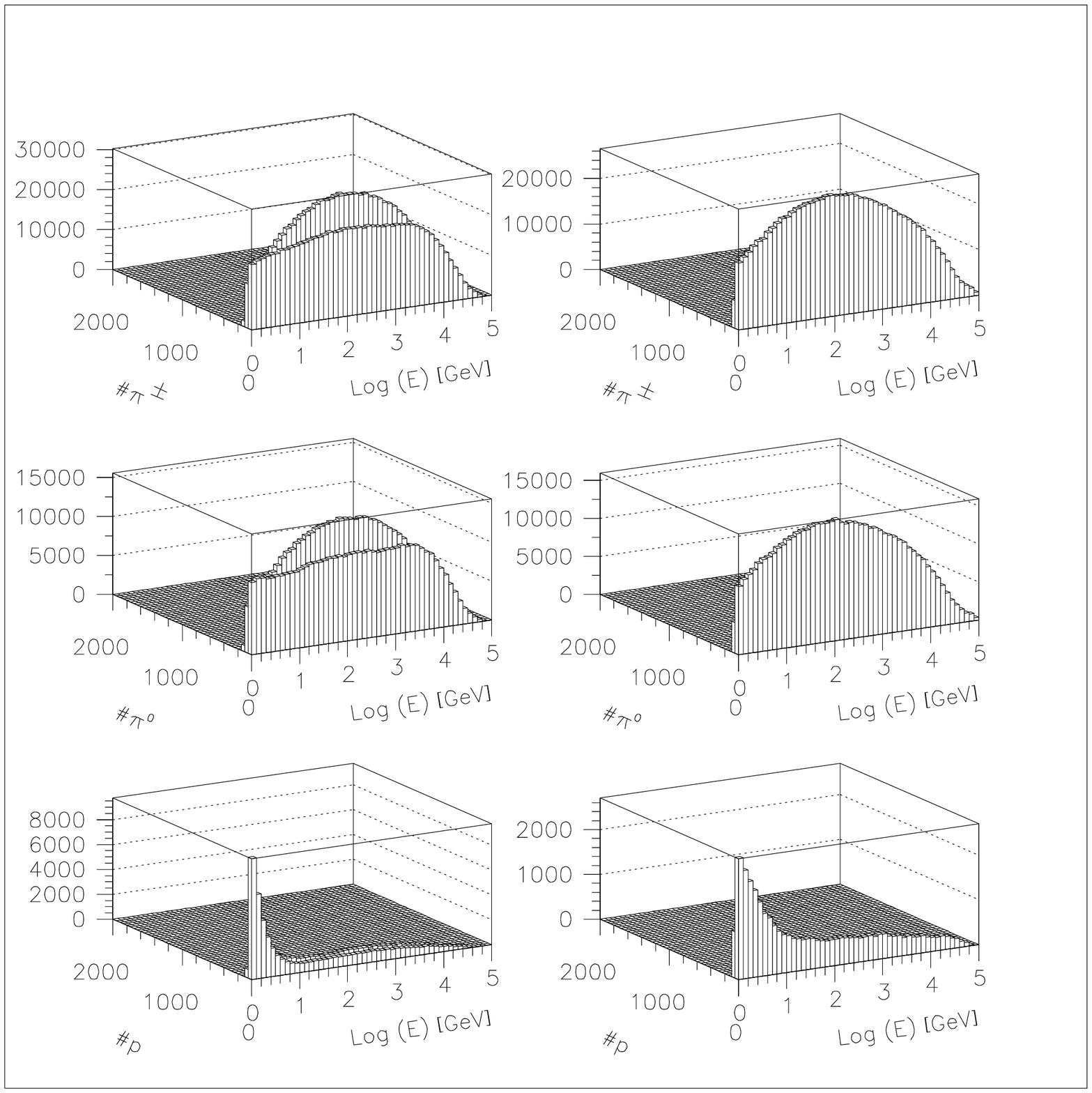,width=15cm,clip=}
\caption{}
\end{center}
\end{figure}

\newpage

\begin{figure}[htpb]
\label{pacota2}
\begin{center}
\epsfig{file=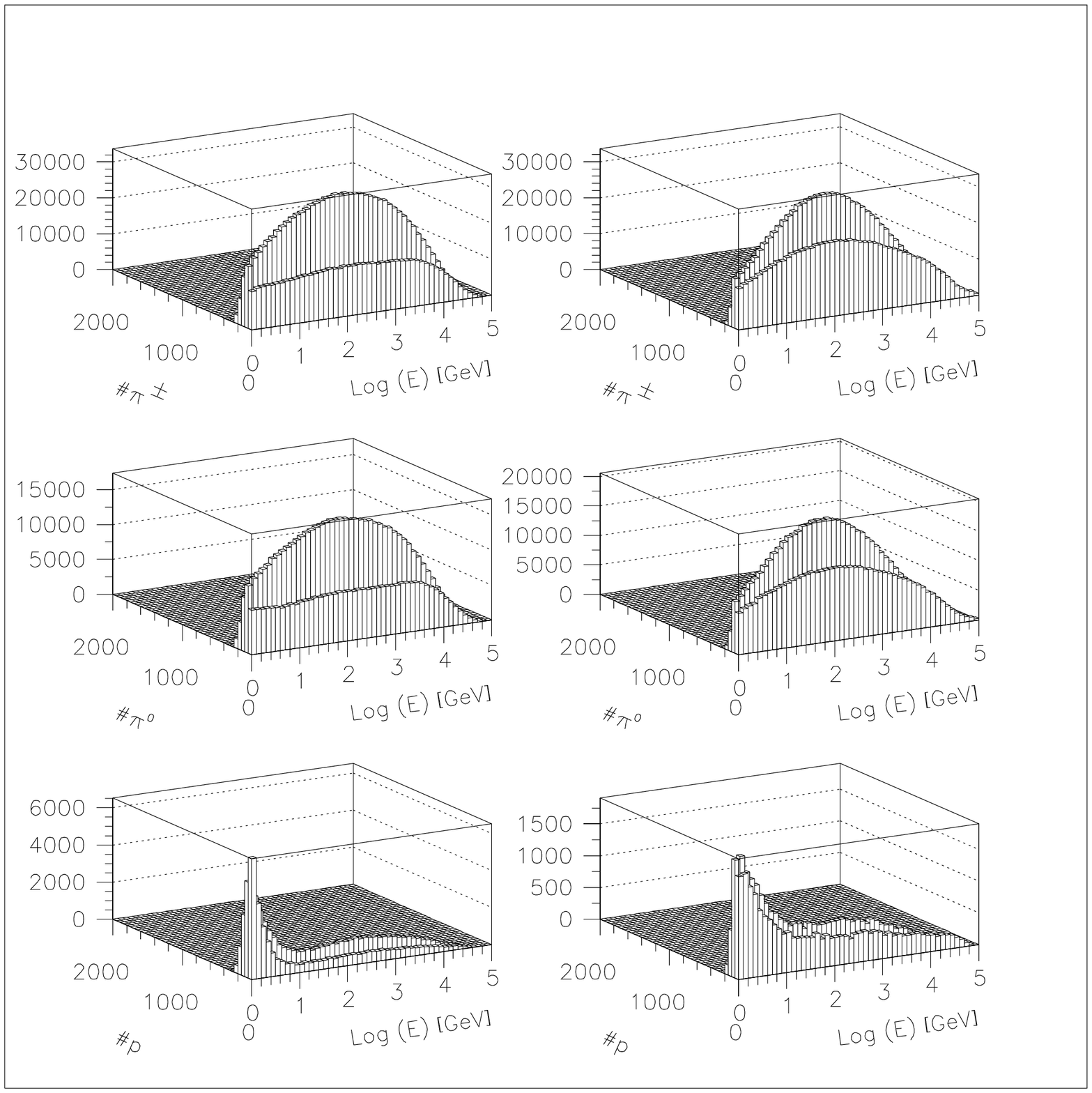,width=15cm,clip=}
\caption{}
\end{center}
\end{figure}

\newpage

\begin{figure}[htpb]
\label{pacota7}
\begin{center}
\epsfig{file=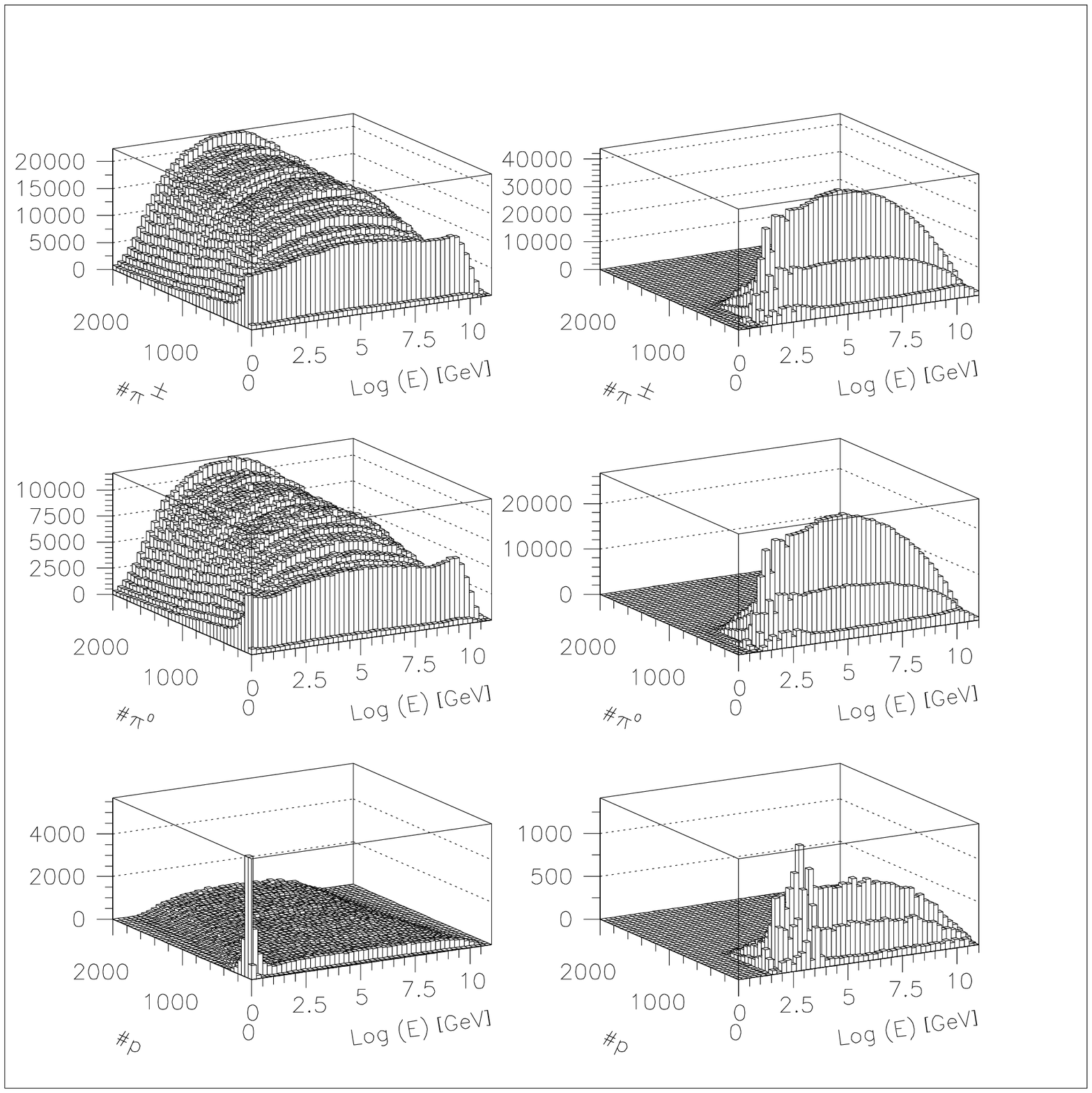,width=15cm,clip=}
\caption{}
\end{center}
\end{figure}

\newpage

\begin{figure}[htpb]
\label{pacota8}
\begin{center}
\epsfig{file=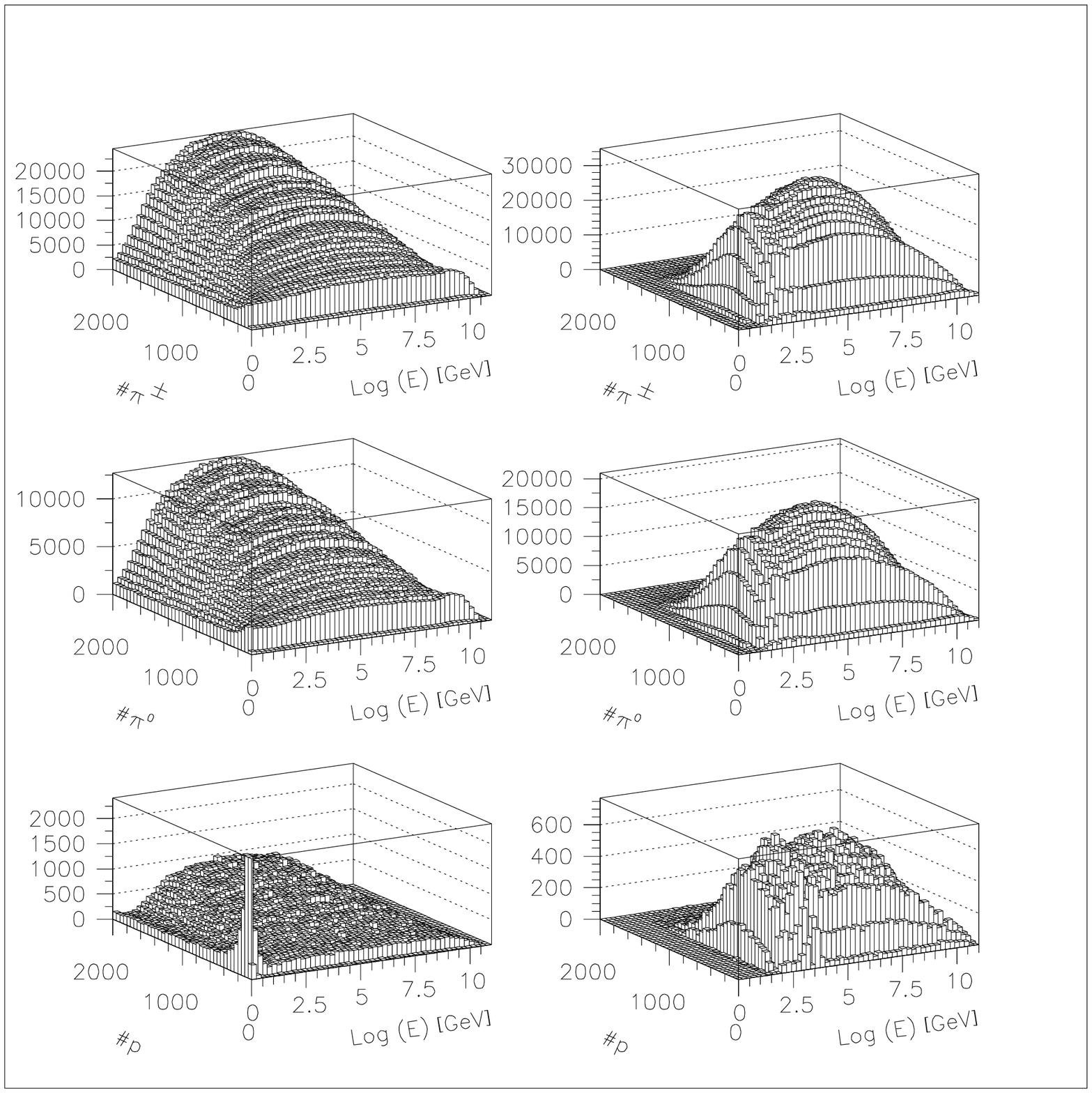,width=15cm,clip=}
\caption{}
\end{center}
\end{figure}

\newpage

\begin{figure}[htpb]
\label{xmax}
\begin{center}
\epsfig{file=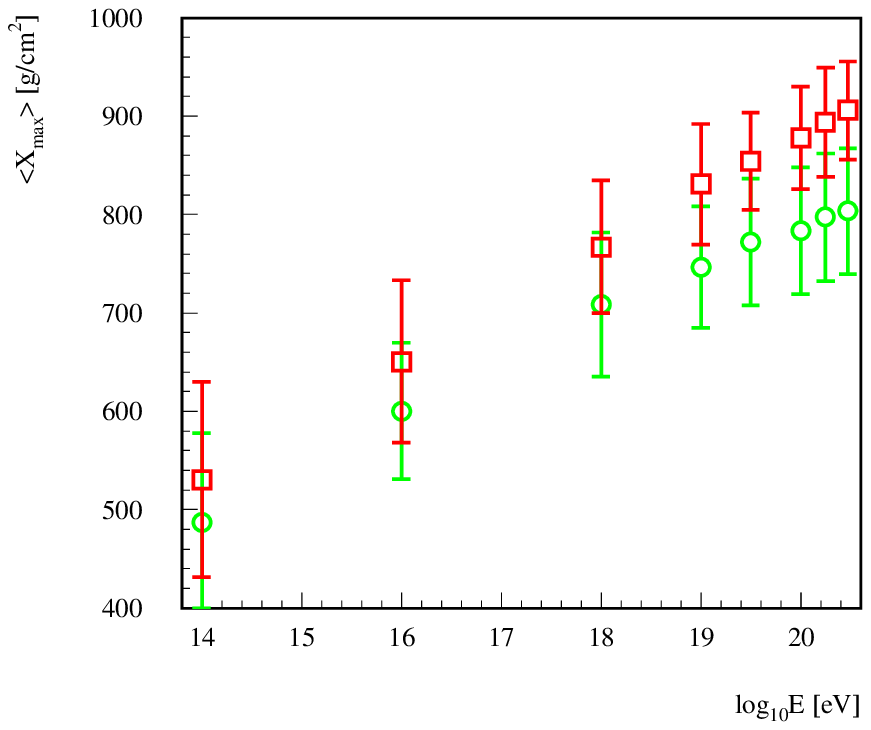,width=15cm,clip=}
\caption{}
\end{center}
\end{figure}

\newpage
\begin{figure}[htpb]
\begin{center}
\label{balloon}
\epsfig{file=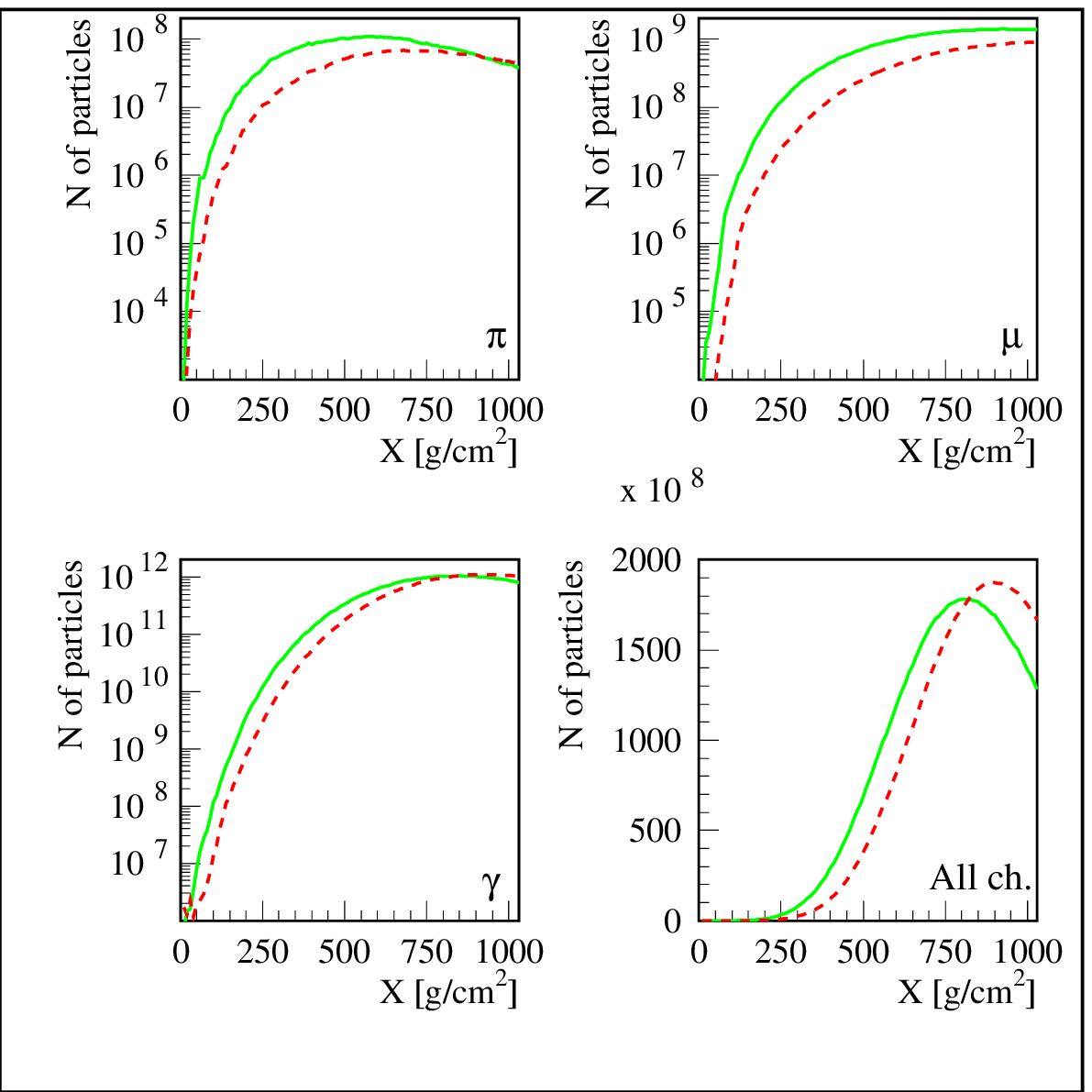,width=15cm,clip=}
\caption{}
\end{center}
\end{figure}

\newpage

\begin{figure}[htpb]
\label{owene}
\begin{center}
\epsfig{file=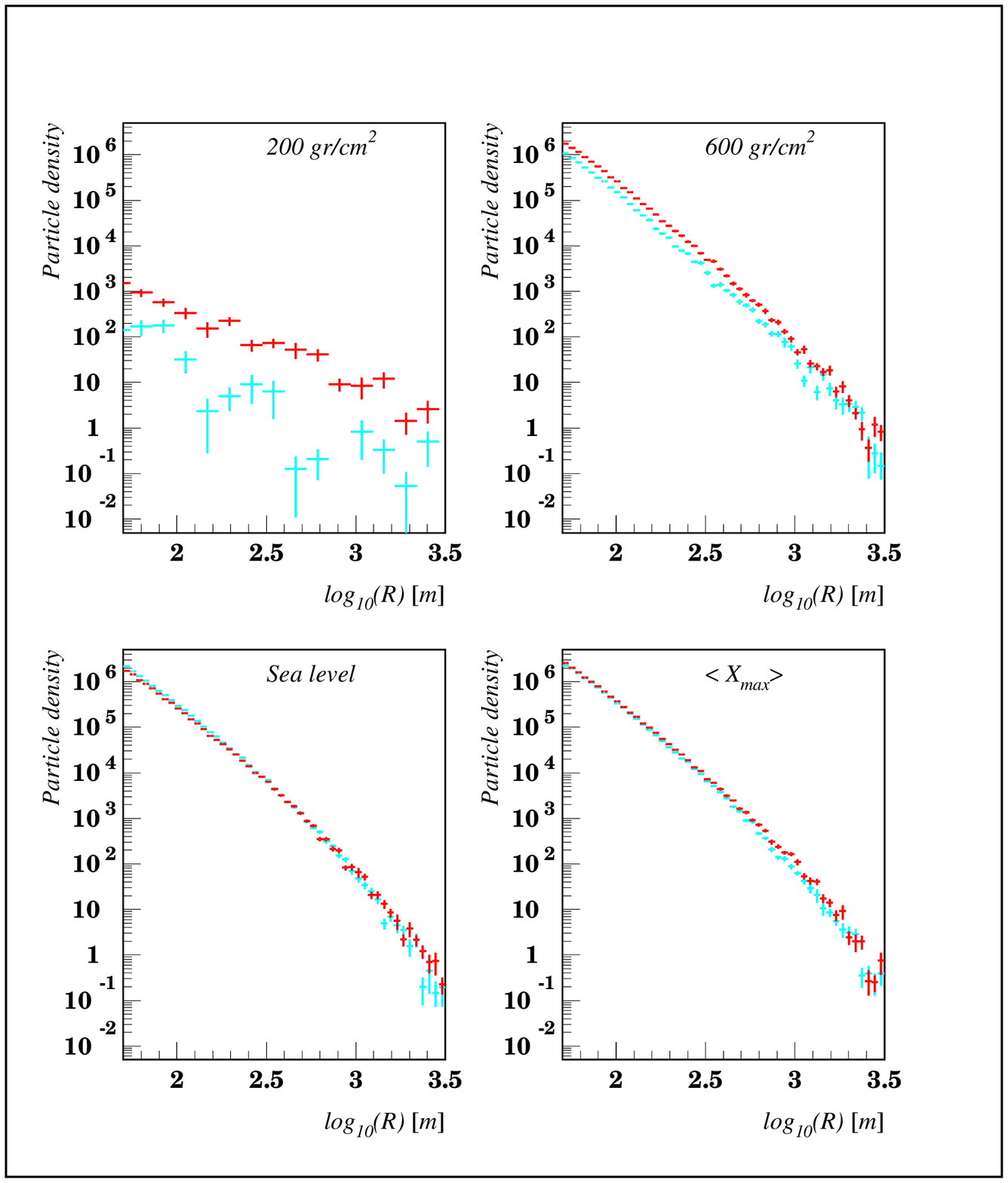,width=15cm,clip=}
\caption{}
\end{center}
\end{figure}

\newpage

\begin{figure}
\begin{center}
\label{owenmu}
\epsfig{file=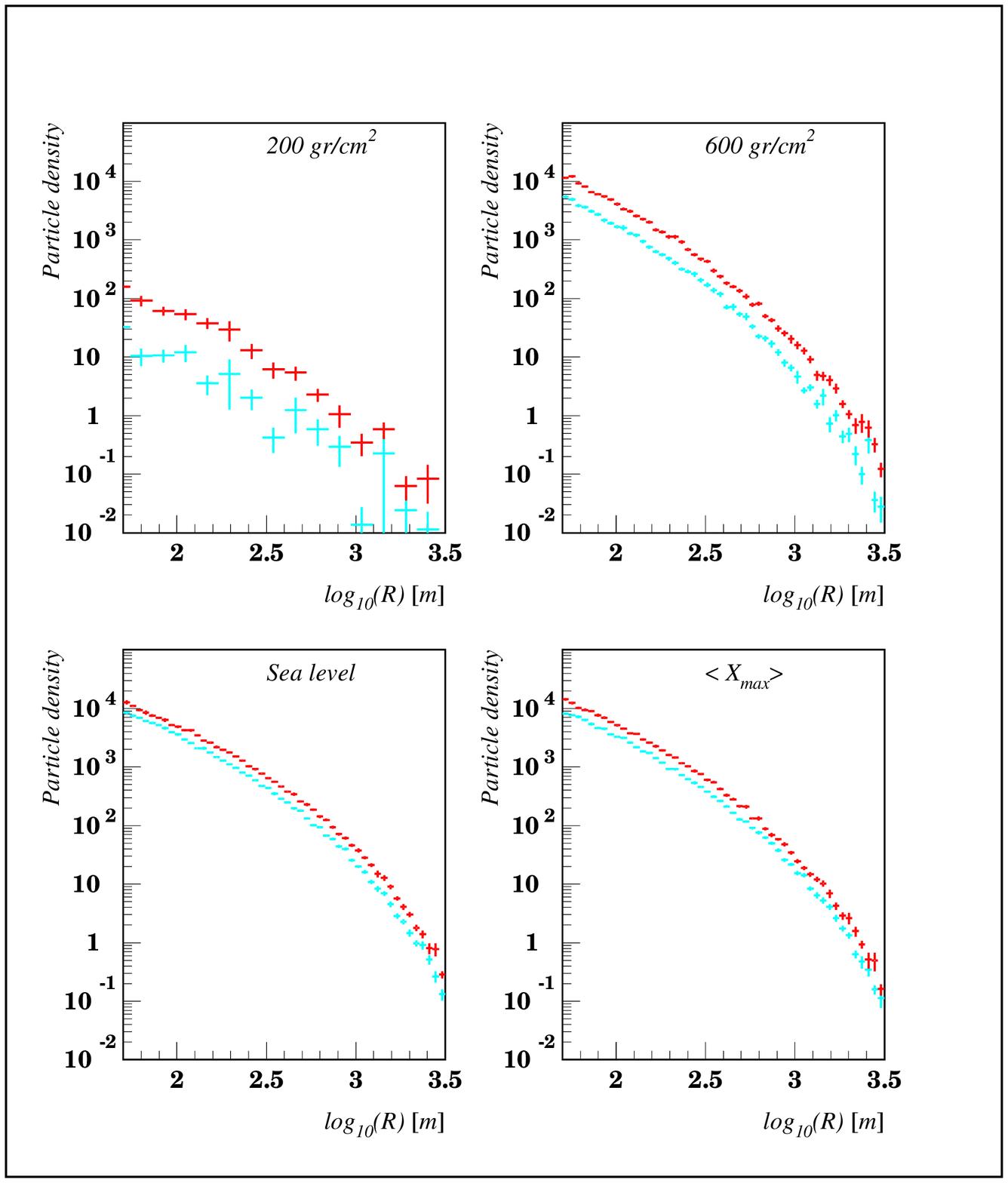,width=15cm,clip=}
\caption{}
\end{center}
\end{figure}

\newpage

\begin{figure}
\label{owenga}
\begin{center}
\epsfig{file=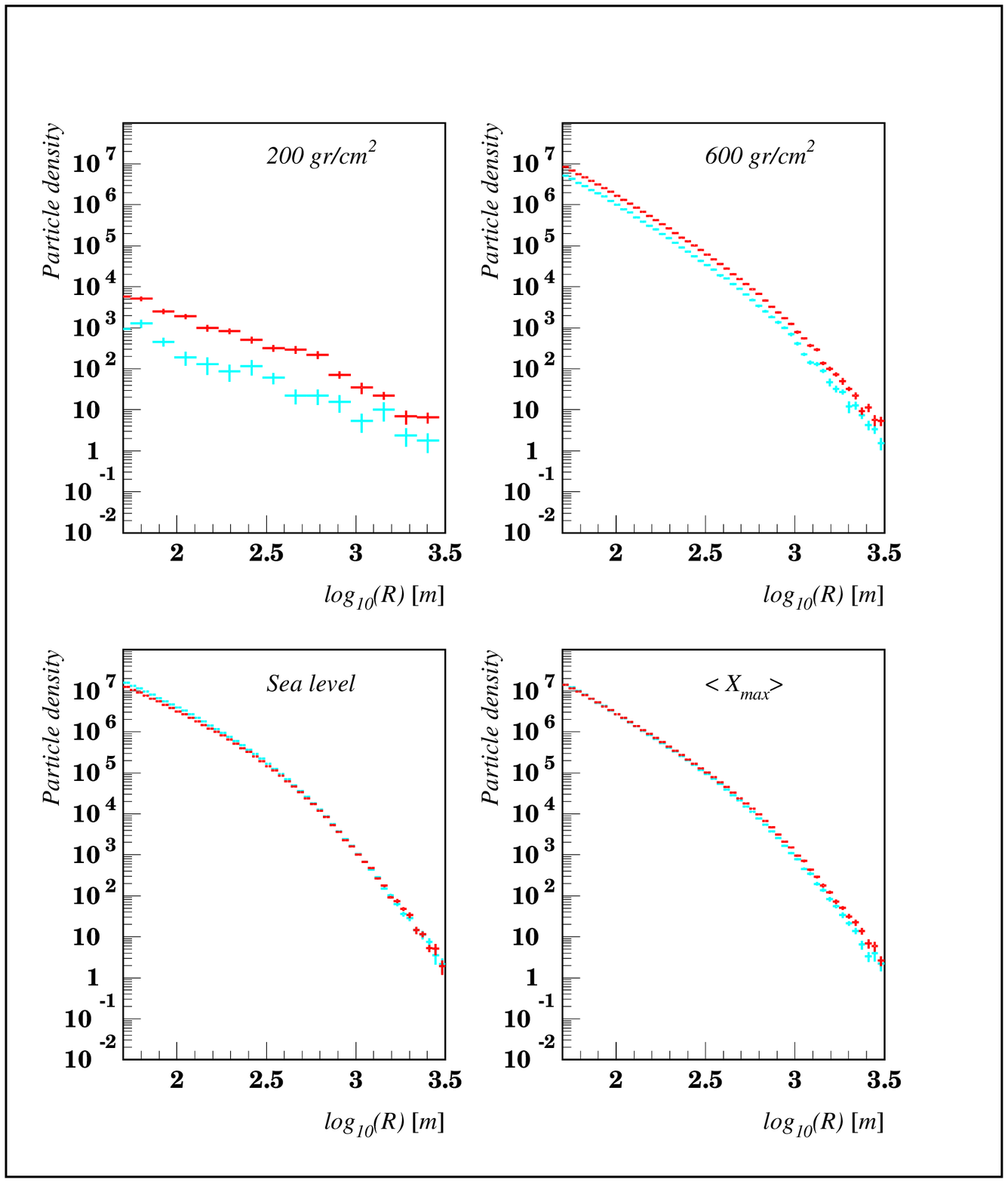,width=15cm,clip=}
\caption{}
\end{center}
\end{figure}

\newpage

\begin{figure}
\label{lenel}
\begin{center}
\epsfig{file=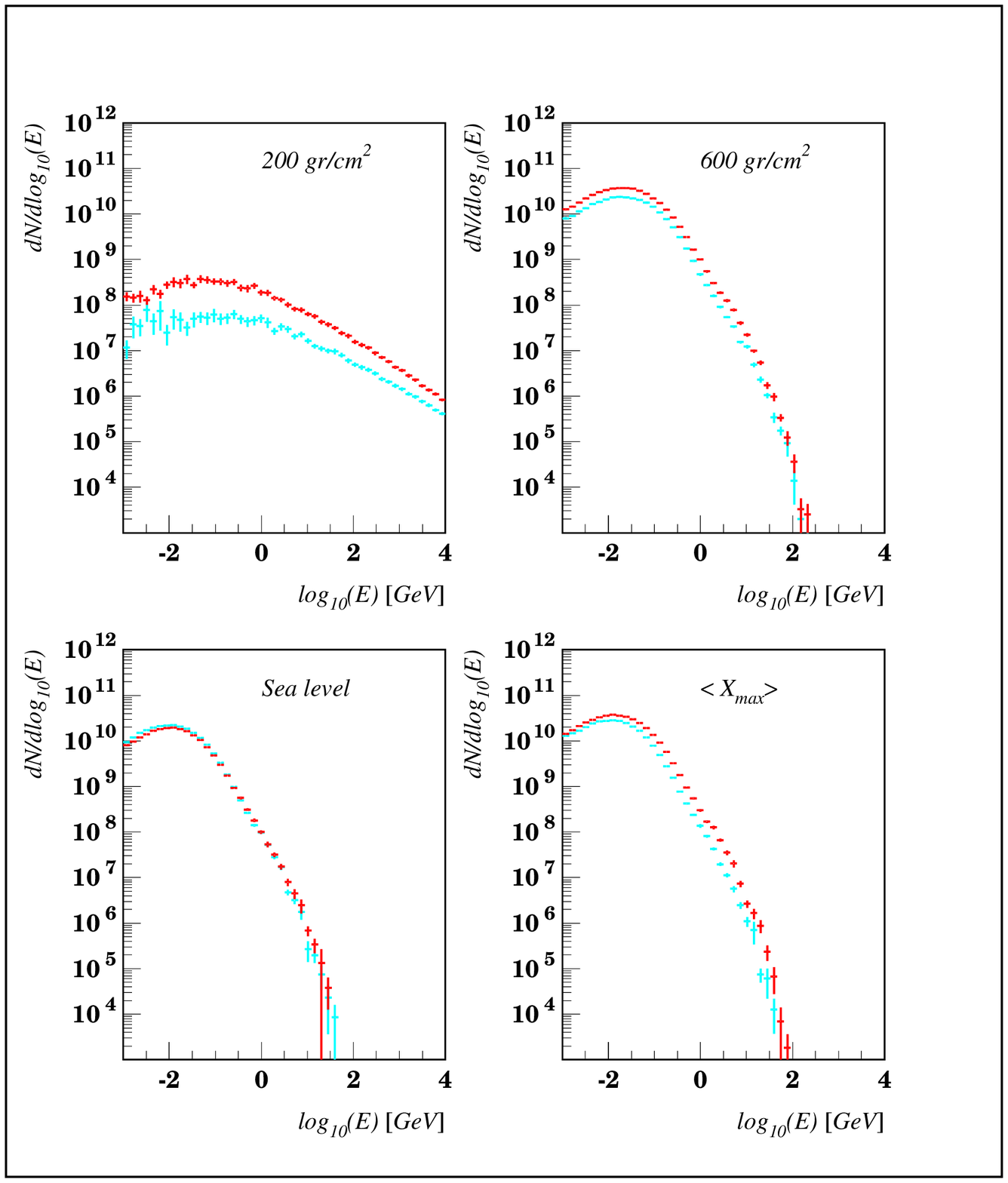,width=15cm,clip=}
\caption{}
\end{center}
\end{figure}

\newpage

\begin{figure}
\label{lenmu}
\begin{center}
\epsfig{file=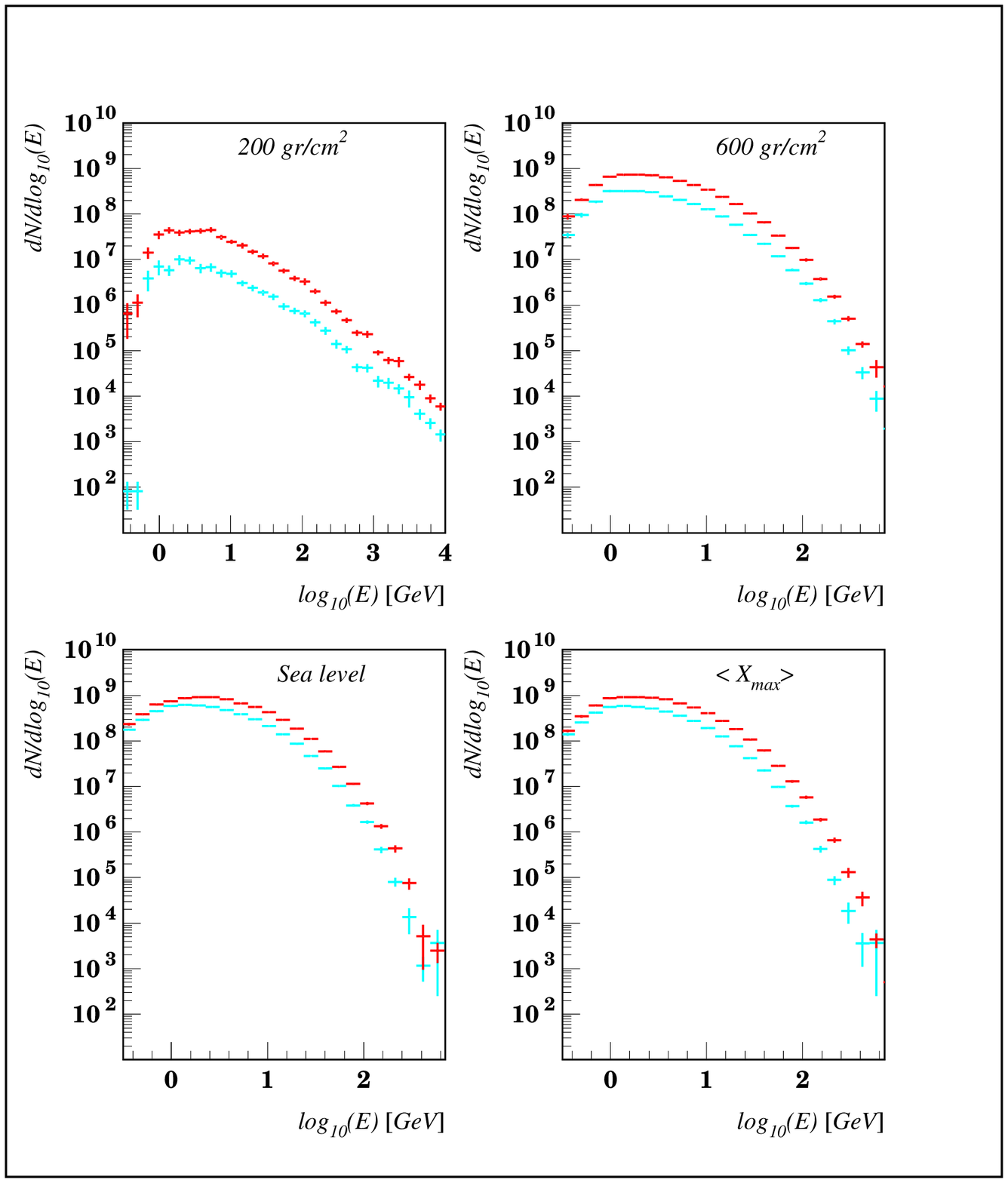,width=15cm,clip=}
\caption{}
\end{center}
\end{figure}

\newpage

\begin{figure}
\label{lenga}
\begin{center}
\epsfig{file=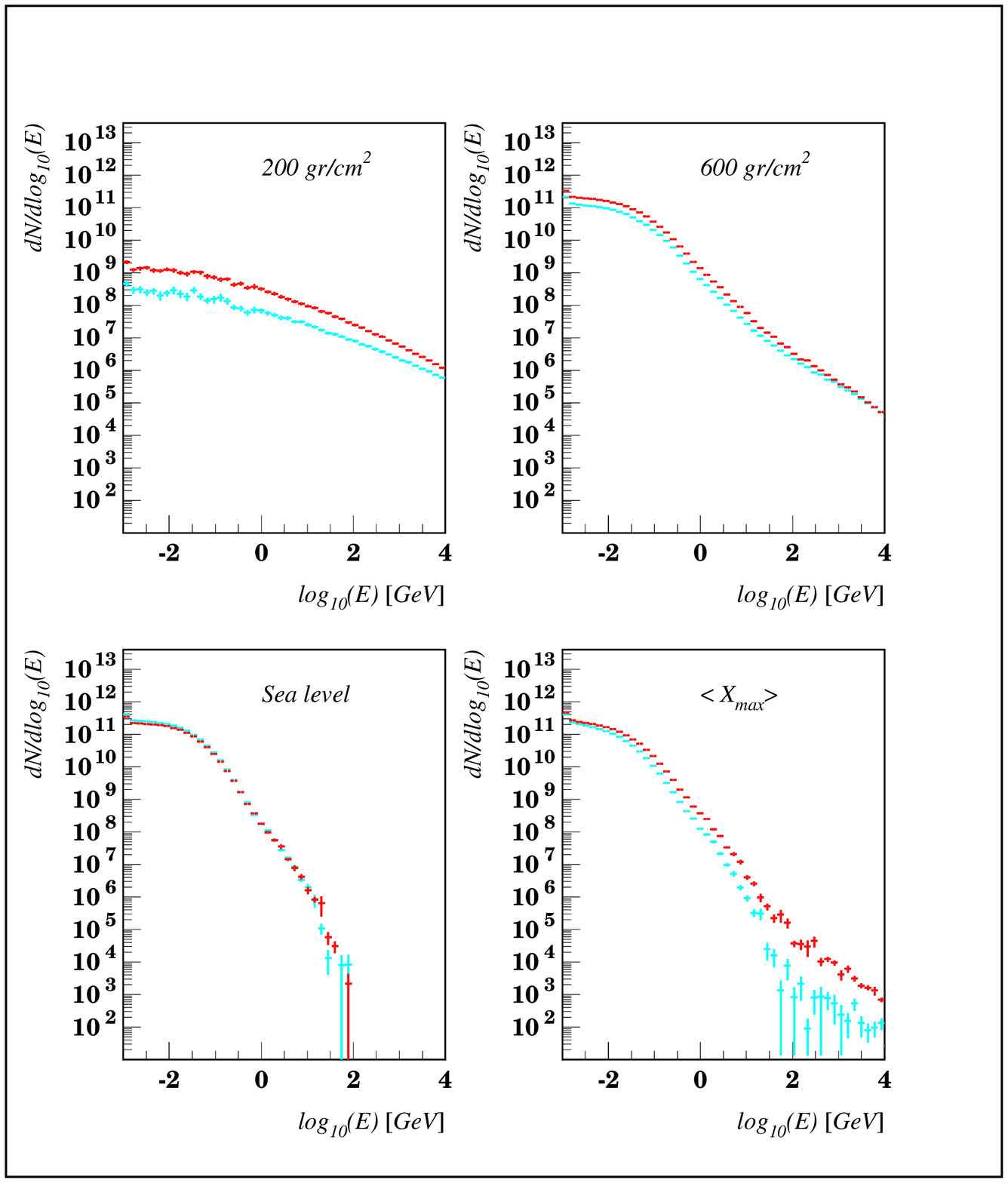,width=15cm,clip=}
\caption{}
\end{center}
\end{figure}


\begin{thebibliography}{99}

\bibitem{Linsley} It is interesting to recall that the first events
with an energy $\sim$ 10$^{20}$ eV were recorded by Volcano Ranch in
the sixties; J. Linsley, Phys. Rev. Lett. {\bf 10}, 146 (1963). These
events together with the ones recently reported (D. J. Bird {\it et
al.} Phys. Rev. Lett. {\bf 71}, 3401 (1993); N. Hayashida {\it et
al.}, Phys. Rev. Lett. {\bf 73}, 3491 (1994); M. Takeda {\it et al.}
Phys. Rev. Lett. {\bf 81}, 1163 (1998)) became a challenge
to current models for the
production of cosmic rays because of the predicted cutoff in the
energy spectrum; K. Greisen, Phys. Rev. Lett. {\bf 16}, 748 (1966); G.
T. Zatsepin and V. A. Kuz'min, Pis'ma Zh. \'Eksp. Teor. Fiz. {\bf 4},
114 (1966) [JETP Lett. {\bf 4}, 78 (1966)].

\bibitem{dpm} A. Capella, U. Sukhatme,
C. I. Tan and J. Tran Thanh Van, Phys. Rep. {\bf 236}, 225 (1994).

\bibitem{qgs} A. B. Kaidalov, Phys. Lett B {\bf 116}, 459 (1982); A.
B. Kaidalov and K. A. Ter-Martirosyan, Phys. Lett. B {\bf 117}, 247
(1982); Yad. Fiz. {\bf 39}, 1545 (1984) [Sov. J. Nucl. Phys.
{\bf 39}, 979 (1984)].

\bibitem{UA1} UA1 Collaboration, C. Albajar et al., Nucl. Phys. B {\bf 309},
405 (1988).

\bibitem{ua1} UA1 Collaboration, G. Arnison et al. Phys. Lett. B
{\bf 118}, 167 (1982).

\bibitem{sybill} R. S. Fletcher, T. K. Gaisser, P. Lipari and T.
Stanev, Phys. Rev. D {\bf 50}, 5710 (1994).

\bibitem{qgsjet} N. N. Kalmykov, S. S. Ostapchenko, A. I. Pavlov,
Nucl. Phys. B (Proc. Supp.) {\bf B52}, 17 (1997).

\bibitem{minijet} T. K. Gaisser and T. Stanev, Phys. Lett. B {\bf
219}, 375 (1989).

\bibitem{durandpi} L. Durand and H. Pi, Phys. Rev. Lett. {\bf 58}, 303 (1987).

\bibitem{kaida} Details on hadron-nucleus interactions as described
by QGSJET are discussed in, A. B. Kaidalov, K. A. Ter-Martirosyan and
Yu. M. Shabel'skii, Yad. Fiz. {\bf 43}, 1282 (1986)
[Sov. J. Nucl. Phys. {\bf 43}, 822 (1986)].

\bibitem{glauber} R. J. Glauber, Nucl. Phys. B {\bf 21}, 135 (1970).

\bibitem{frichter} G. M. Frichter, T. K. Gaisser and T. Stanev, Phys. Rev. D
{\bf 56}, 3135 (1997).


\bibitem{Knapp} A comparison of
hadronic interaction models used in air shower
simulations have been already performed
by the group of Karlsruhe with the code CORSIKA. The analysis was based
on proton and iron induced showers with primaries energies ranging
from 10$^{14}$ to $10^{15}$ eV. See, J. Knapp, D. Heck and G. Schatz,
Nucl. Phys. B (Proc. Suppl.) {\bf 52B}, 136 (1997);
Report FZKA 5828 (1996); D. Heck, J. Knapp, and G. Schatz,
Nucl. Phys. B (Proc. Suppl.) {\bf 52B}, 139 (1997).


\bibitem{sergio} S. J. Sciutto, {\it AIRES: A System for Air Shower
Simulations}, Auger technical note GAP-98-032 (1998); available
electronically from: http://www-td-auger.fnal.gov:82.


\bibitem{mocca} Most of the electromagnetic algorithms are based on
the well known MOCCA simulation program by A. M. Hillas, Nucl. Phys. B
(Proc. Suppl.) {\bf 52}, 29 (1997).

\bibitem{carroll} A. S. Carroll {\it et al.}, Phys. Lett. B {\bf 80},
319 (1979).

\bibitem{roberts} T. J. Roberts {\it et al.}, Nucl. Phys. B {\bf
159}, 56 (1979).

\bibitem{hara} T. Hara {\it et al}, Phys. Rev. Lett. {\bf 50}, 2058 (1983).
It is important to stress that while computing these data an energy
dependence on
$\sigma_{p{\rm - air}}$ represented as $290\, E_{\rm lab}^{0.06 \pm 0.01}$
was taken into account. This parametrization was obtained with the
assumption that there is no significant break of Feynman scaling in
the fragmentation region and that the multiplicity increases as $\ln^2 s$.
The value of $\sigma_{p{\rm - air}}$ is expected to become a little
smaller if there is a significant breakdown of scalling in the fragmentation
region. See, \cite{honda}.

\bibitem{honda} M. Honda {\it et al.}, Phys. Rev. Lett. {\bf 70}, 525
(1993).


\bibitem{baltrusaitis} R. M. Baltrusaitis {\it et al.}, Phys. Rev. Lett.
{\bf 52}, 1380 (1984).

\bibitem{Tom} T. K. Gaisser, private communication; see also SIBYLL source
code.


\bibitem{hillas} A. M. Hillas, in {\it Proc. of the 16$^{\rm th}$
International Cosmic Ray Conference}, Tokyo, Japan, 1979
(University of Tokyo, Tokyo, 1979), Vol.8,p.7.; updated in,
{\it Proc. of the 17$^{\rm th}$ International Cosmic Ray Conference},
Paris, France, 1981 (CEN, Saclay, 1981), Vol.8,p.183.



\bibitem{pryke} C. Pryke and L. Voyvodick, (to appear in {\it Proc. of
the 10$^{\rm th}$ International Symposium on Very High Energy Cosmic Ray
Interactions}, July 12-17, 1998, LNGS, Assergi, Italy). Auger technical
note GAP-98-052 (1998); available electronically
from: http://www-td-auger.fnal.gov:82.

\bibitem{agasa} M. Nagano, Nucl. Phys. B (Proc. Suppl.)
{\bf 52}, 71 (1997).

\bibitem{auger} The Auger Collaboration, Pierre Auger Project Design
Report, Fermi National Accelerator Laboratory, available
electronically from, http://www-td-auger-fnal.gov:82.

\bibitem{owl} J. F. Ormes {\it et al.}, in {\it Proc. of the $25^{\rm th}$
International Cosmic Ray Conference}, (Durban, S.A.) M. S. Potgieter,
B. C. Raubenheimer and D. J. van der Walt Eds., 1997, Vol.5, p.273.

\end{thebibliography}
\end{document}